# Increasing FPS for single board computers and embedded computers in 2021 (Jetson nano and YOVOv4-tiny).
## Practice and review

Ildar Rakhmatuiln[a]*Department of Power Plants Networks and Systems, South Ural State University, Chelyabinsk city, Russia*

ildar.o2010@yandex.ru

**Abstract**

This manuscript provides a review of methods for increasing the frame per second of single-board computers. The main emphasis is on the Jetson family of single-board computers from Nvidia Company, due to the possibility of using a graphical interface for calculations. But taking into account the popular low-cost segment of single-board computers as RaspberryPI family, BananaPI, OrangePI, etc., we also provided an overview of methods for increasing the frame per second without using a Graphics Processing Unit. We considered frameworks, software development kit, and various libraries that can be used in the process of increasing the frame per second in single-board computers. Finally, we tested the presented methods for the YOLOv4-tiny model with a custom dataset on the Jetson nano and presented the results in the table.

**Nomenclature**

| | |
|---|---|
| CNNs | Convolutional Neural Networks |
| DCNNs | Deep Convolutional Neural Networks |
| SBC | Single Board Computer |
| GPU | Graphics Processing Unit |
| SDK | Software development kit |
| FPS | Frame per second |
| CUDA | Compute Unified Device Architecture |

Our Dataset for train YOLOv4-tiny - https://www.kaggle.com/ildaron/tracking-a-cockroach-at-home

## 1. Introduction

Machine vision is the fastest growing trend in artificial intelligence in the last few years. Today, machine vision is not only studied by IT professionals, but also used in a wide variety of industries by specialists from different fields. For example, RaspberryPI (https://www.raspberrypi.org/), which is the flagship in the field of SBC, is quite often used in everyday situations. Seelam et al. (3, 2021) used this SBC to monitor classroom attendance. Tagnithammou et al. (5, 2021) used RaspberryPI for more specific task to detect an emotional state in a mobile vehicle. Rakhmatulin (19, 2021; 20, 2021) used Raspberry Pi for kill mosquito and insect by laser. Krishna et al. (4, 2021) dispel the myth of the unreliability of SBCs and used RaspberryPI to protect against network intrusions using the example of Snort IDS (intrusion detection system). Most often, due to limited computing power in RaspberryPI for object detection tasks in real-time researchers used Haar cascades (18, 2021). Haar traits are digital image traits used in pattern recognition. They owe their name to an intuitive resemblance to Haar wavelets. But Haar cascades have several disadvantages compared to DCNNs, as detailed by Raj et al. (6, 2020), Kaur et al (7, 2021). The main one of which is the impossibility of detecting objects when changing the monitoring angle. As a result, the area of use of Haar cascades for real-time object detection began to decrease with the growing popularity of CNNs.

The use of GPUs in the calculations of neural networks made a breakthrough in the field of machine vision and allowed the use of neural networks for a wide range of researchers. For SBCs in real-time tasks, the use of CUDNN became possible largely due to CUDA. CUDA is a software and hardware parallel computing architecture that can significantly increase computing performance using Nvidia GPUs. This makes it possible to use SBC from Nvidia in more difficult machine vision tasks. For example, Zhao et al. (10, 2021) used Jetson Nano to accurately sort high quality soybean seeds. Zhao et al. (8, 2021) also with the Jetson Nano board implemented a system for recognizing defects of the entire surface of soybean seeds in real-time based on deep learning. Wanshui, G. et al. [9] used the more expensive version of NVIDIA - Jetson TX2 for real-time adaptive stereo depth estimation.

The above examples clearly show that machine vision has practical applications, therefore, the implementation of this technology requires special conditions for hardware, the main of which is the low-cost of equipment and computing power - in the context of machine vision, we can say FPS.

The area of SBC implementation is large, and its application is only constrained by the fact that the depth of neural networks increases faster than the hardware part becomes cheaper. But fortunately, there are frameworks and libraries, and other various tools designed to speed up the work of SBCs. Frameworks, libraries for machine vision are updated with a frequency that is not inherent in any other field of science. Therefore, it is very difficult to know all the available opportunities for raising FPS on the SBC. As a result of which the purpose of this manuscript is to conduct a detailed analysis of the state of this issue for 2021 year. There are several works in which the authors gave an overview of methods aimed at increasing FPS in SBC, but the relevance of these works is lost after only 1-2 years. Beckhauser et al. (2, 2016) in 2016, after the analysis, concluded that the Raspberry PI3 was the best solution in the SBC market. Basford et al. (1, 2020) came to a similar conclusion four years later when compared at the Raspberry Pi 3 Model B, Raspberry Pi 3 Model B +, and Odroid C2. Mittal et al (11, 2019) presented a paper that is as close as possible to our paper in terms of topics. Their review presents the evaluation and optimization of neural network applications on the Jetson platform. The authors reviewed both hardware and algorithmic optimizations performed to run neural network algorithms. But the authors considered only the Jetson SBC family, and over the past 2 years from the date of publication, a lot of new software has been released.

## 2. Methods for increasing FPS on SBC

In this chapter, we will look at the most popular methods for increasing FPS on SBC. But first, to fully understand the following information, we considered a little theory:

- **Int8, Int16:** The int 8, 16 value is related to storage capacity. Variables with int8 can range from -128 to 127, and int16 from -32 768 to 32 767. The wider the range, the more SBC memory needs to be allocated for working with variables.
- **FP16, FP32:** Machine learning models in most frameworks are trained with 32-bit precision. In the case of FP16, the accuracy of the trained model is reduced by 16 bits. The advantages of FP16 are that it improves speed (TFLOPS) and performance while reducing neural network memory usage. But FP16s must be converted from 32-bit floating point numbers before working with them, which reduces the efficiency of the neural network.

We used Jetson nano SBC to test the proposed methods. To implement the neural network, we used Yolov4-tiny to search for 1 object in real-time video with a resolution of 416x416 pixels from a Sony IMX219 camera. We choose Yolo because this neural network has several advantages that clear were described by Bochkovskiy, A. et al. (16, 2020). In addition, we would like to add that is possible to use nVidia Transfer Learning Toolkit to train YOLOv4, since this Toolkit has a built-in native conversion to nVidia TensorRT (https://docs.nvidia.com/metropolis/TLT/tlt-user-guide/text/object_detection/yolo_v4.html). Also, support for the classic YOLO is available in the most popular frameworks (TF/PT/TRT/OpenCV/OpenVino/ONNX-runtime/) and works on popular hardware models (CPU/GPU nVidia/AMD /NPU/FPGA/) https://github.com/AlexeyAB/darknet#yolo-v4-in-other-frameworks

### 2.1. Keras

To begin, we implemented YoloV4-tiny on Keras (https://keras.io/). This open neural network library is one of the most popular among researchers due to its efficiency and comprehensibility. Keras is a minimalist Python-based library that can run on top of TensoFlow, Theano, or CNTK. Keras is not as powerful as TensorFlow and provides fewer options for managing the network connection, which can be a serious limitation.

As a result, we used the following git:

https://github.com/jkjung-avt/tensorrt_demos

and got 4-5 fps. Quite low speed, which will not allow working in real time with Keras.

### 2.2. cuDNN

After that we implemented YoloV4-tiny on the Darknet framework (https://pjreddie.com/darknet/).

Darknet - An open framework for application use of a CNN, suitable, for example, for classifying photographs or highlighting objects in images in real-time. Darknet is written in C using CUDA with the ability to use cuDNN. cuDNN (https://developer.nvidia.com/cudnn) is a GPU-accelerated library of primitives for ultra-precise neural networks. This library can be embedded in high-level machine learning frameworks. CuDNN adds support for storing 16-bit floating point data in GPU memory, doubling the storage capacity, and optimizing memory bandwidth usage. That will allow you to train more deep neural networks.

As a result, we used the following git:

https://github.com/AlexeyAB/darknet

and got 12-14 fps, which is already can be acceptable for some detection tasks in real-time.

**2.3. NVIDIA TensorRT** (https://developer.nvidia.com/tensorrt)

NVIDIA released TensorRT SDK for high-performance deep learning output. This SDK includes a deep learning inference optimizer and a runtime that provides low latency and high throughput for deep learning inference applications. According to the manufacturers, TensorRT-based applications run up to 40x faster during inference than CPU-only platforms. With TensorRT, it is possible to optimize neural network models trained on all major platforms, calibrate for lower accuracy with high accuracy, and deploy to hyperscale data centers. TensorRT provides INT8 and FP16 optimizations for production deployments of deep learning inference applications. TensorRT is also integrated with application-specific SDKs such as NVIDIA DeepStream, Jarvis, Merlin, Maxine, and Broadcast Engine and is integrated with Matlab and ONNX. ONNX Runtime Python package and Docker container for the NVIDIA Jetson platform, now available on the Jetson Zoo.

As a result, we used the following git:

https://github.com/jkjung-avt/tensorrt_demos

and got 25 fps, which is already enough for most tasks.

**2.4. DeepStream SDK** (https://developer.nvidia.com/deepstream-sdk)

NVIDIA released the DeepStream SDK. Optimized for NVIDIA Jetson platforms, this SDK is built on top of the GStreamer framework (https://gstreamer.freedesktop.org/). This SDK provides an end-to-end service for converting raw streaming data into analytic data. The DeepStream SDK also enables the simultaneous use of multiple neural networks to process each individual video stream. As a result, we used the following git:

https://github.com/marcoslucianops/DeepStream-Yolo

and got about the similar FPS as for TensorRT - 25 fps.

**2.5. tkDNN**

tkDNN is a deep neural network library built with cuDNN and tensorRT primitives specifically designed to run on NVIDIA Jetson boards. The authors of this library are Verucchi et al. (12, 2020) published a work in which they tested this library on TX1, TX2, AGX Xavier.

As a result, we used the following git:

https://github.com/ceccocats/tkDNN

and got 30-35 fps, which is the best solution now.

Table 1 shows the results of using the proposed above methods for Yolov4-tiny on a Jetson Nano for tracking a single object. We deliberately overlooked the different parameters of neural networks, such as complexity of settings, required memory, and hardware power consumption, and simplified the table as much as possible to reflect the topic of this study, namely FPS.

Table 1. FPS figures for the Yolov4-tiny model when running on the Jetson Nano

| **Методы** | Keras | Darknet | Darknet Tensor RT | Darket DeepStream | tkDNN |
|---|---|---|---|---|---|
| **FPS** | 4-5 | 12-15 | 24-27 | 23-26 | 30-35 |

## 3. Other recommendations for increase FPS

### 3.1. Amazon Web Services (AWS) IoT Greengrass (https://aws.amazon.com/)

On Linux devices, it is possible to use out-of-the-box solutions such as the AWS IoT Greengrass platform. The platform effectively extends AWS capabilities to edge devices, allowing them to work locally with the data they create while using the cloud to manage, analyze, and securely store data.

### 3.2. Google Colab (https://colab.research.google.com/)

Google Colab is a service like Jupyter-Notebook that offers free access to a GPU. Colab has been updated to the new NVIDIA T4 GPUs, which made it possible to experiment with RAPIDS at Colab. Rapids (https://developer.nvidia.com/rapids) is a set of software libraries based on CUDA-X AI that enables end-to-end data and analytics pipelines to run entirely on GPUs. Washburn et al. (17, 2021) showed that the RAPIDS CUDA machine learning algorithm works entirely on the GPU architecture, and increases the speed of detecting bright formations for the diagnosis of diabetic retinopathy.

### 3.3. OpenVINO, Intel (https://docs.openvinotoolkit.org/latest/index.html)

OpenVINO is an open-source, free toolkit that helps data developers accelerate the development of high-performance solutions for use in a variety of video systems. It is possible to use OpenCV compiled with CUDA and OpenVINO. As a result, the speed on the GPU is like on TensorRT. https://github.com/AlexeyAB/darknet#geforce-rtx-2080-ti

### 3.4. Other common methods

**CuBLAS** (https://developer.nvidia.com/cublas) - it is a library for basic matrix calculations. The cuBLAS library provides a GPU-accelerated implementation of basic linear algebra routines (BLAS). The cuBLAS library is included in both the NVIDIA HPC SDK and the CUDA Toolkit. Suita, et al. (13, 2020) described efficient implementations of a convolutional pool in a GPU in which join follows multiple convolutions. For this, the convolution exchange is converted to matrix multiplication, which can be computed with cuBLAS very efficiently.

**Library cuFFT** (http://docs.nvidia.com/cuda/cufft/index.html) is designed to deliver high performance on NVIDIA GPUs. Galletti, A. et al. (14, 2017) uses the NVIDIA cuFFT library to implement a noise reduction algorithm to remove noise in the frequency domain.

Set of tools **NVIDIA VisionWorks** (https://developer.nvidia.com/embedded/visionworks) - is a computer vision (CV) and image processing software development package. VisionWorks implements and extends the Khronos OpenVX standard (https://www.khronos.org/openvx/) and is optimized for CUDA-enabled GPUs, allowing developers to implement CV applications on a scalable and flexible platform. Qasaimeh, et al. (15, 2021) successfully tested neural network inference accelerators on embedded platforms using VisionWorks.

It is also worth considering various auxiliary tools such as the NVIDIA **Vision Programming Interface (VPI)**. (https://docs.nvidia.com/vpi/index.html ). VPI is a software library that implements computer vision and image processing algorithms on several computing hardware platforms available in NVIDIA embedded. And **Xilinx Deep Learning Processor Unit (DPU) (**https://www.xilinx.com/products/intellectual-property/dpu.html**)** - it is a programmable engine designed for a convolutional neural network.

The following devices are of particular interest**. Coral USB Accelerator (**https://coral.ai/products/accelerator/**)** enables SBCs to leverage the power of AI applications by allowing them to inference pre-trained Tensorflow Lite models locally on their own hardware. Also, a similar model from Intel Intel Movidius Neural Compute Stick (https://software.intel.com/content/www/us/en/develop/hardware/neural-compute-stick.html) of a specialized computer in the USB form factor dongle

Microsoft and NVIDIA jointly created, tested and published the **ONNX Runtime Python package (**https://www.onnxruntime.ai/python/tutorial.html**)** and Docker container for the NVIDIA Jetson platform, now available at the Jetson Zoo (https://elinux.org/Jetson_Zoo). Today's release of ONNX Runtime for Jetson extends the performance and portability benefits of ONNX Runtime to Jetson AI edge systems.

**Conclusion and Discussing**

In this manuscript, we have reviewed the most popular and proven solutions for increasing FPS on SBC. We can conclude that today, deep neural networks can be allowed to run on low-cost SBCs (in our case, Jetson Nano) with an acceptable FPS (FPS - 35, YoloV4-tiny, Jetson Nano) due to the use of the described tools. In this paper, we considered only FPS, guided by the fact that in practice this is the most important criterion and did not cover issues such as energy consumption, size, etc., which, if necessary, can be solved, by financial methods. We can conclude that now the market is largely occupied by Nvidia's company, and if the hardware from Nvidia somehow consolidated its position and new models enter the market every few years, then tools for increasing their productivity appear much more often, both from Nvidia and independent researchers. As a result, it is very difficult for specialists who use machine vision in their projects to be aware of all the new frameworks for the single-board computers market.

In the future, it is worth considering in more detail the tools from the Intel Company and its methods aimed at increasing the FPS. There are also dozens of articles in which private methods are presented for accelerating specific models of neural networks, or a specific type of hardware. It is advisable in the future to make review such papers and create a visual classification so that any researcher can easily find the conditions of interest to him for increasing the FPS for his case.

**Conflicts of Interest**: None
**Funding**: None
**Ethical Approval**: Not required